\documentclass[conference]{IEEEtran}
\usepackage[pscoord]{eso-pic}
\IEEEoverridecommandlockouts
\usepackage{graphicx}
\usepackage{cite}
\usepackage{comment}
\usepackage[numbers]{natbib}
\usepackage{caption}
\usepackage{titlesec}

\usepackage{float}
\usepackage{amsmath,amssymb,amsfonts}
\usepackage{mathtools} 
\usepackage{mathtools}
\usepackage{paralist}
\usepackage{booktabs}
\usepackage{algorithmic}
\usepackage{multirow}
\usepackage{tabularx}
\usepackage{tikz}
\usepackage{pgfplots}
\usepackage[caption=false,font=footnotesize]{subfig}
\pgfplotsset{compat=1.18}
\usepackage{textcomp}
\usepackage{subcaption}
\usepackage{todonotes}
\usepackage{xcolor}
\usepackage[font=footnotesize,skip=3pt]{caption}

\titlespacing{\subsection}{0pt}{0.6ex}{0.4ex}

\def\BibTeX{{\rm B\kern-.05em{\sc i\kern-.025em b}\kern-.08em
    T\kern-.1667em\lower.7ex\hbox{E}\kern-.125emX}}

\usepackage{enumitem}
\setlist{noitemsep, topsep=2pt, parsep=2pt, partopsep=0pt}

\begin{document}

\bstctlcite{BSTcontrol}

\title{Exploiting the Alternatives: Coordinated Learning via Hierarchical RL for Dynamic VNEAP\thanks{This work is supported by MUSMET project funded by the
EIC Pathfinder Open scheme of the European Commission
(grant agreement n. 101184379), and has received funding
from the Swiss State Secretariat for Education, Research and
Innovation (SERI). }
}

\author{  
\IEEEauthorblockN{
    Ali Al Housseini\IEEEauthorrefmark{1},
    Cristina Rottondi\IEEEauthorrefmark{2},
    Sebastian Troia\IEEEauthorrefmark{3},
    Omran Ayoub\IEEEauthorrefmark{1}\\
    }

    \IEEEauthorblockA{\IEEEauthorrefmark{1}University of Applied Sciences and Arts of Southern Switzerland, Lugano, Switzerland}
    \IEEEauthorblockA{
    \IEEEauthorrefmark{2}Politecnico di Torino, Turin, Italy;
    \IEEEauthorrefmark{3}Politecnico di Milano, Milan, Italy
}
  }
      

\maketitle
\begin{abstract}
Virtual Network Embedding (VNE) is a key enabler of network slicing, yet most formulations assume that each Virtual Network Request (VNR) has a fixed topology. Recently, VNE with Alternatives (VNEAP) was introduced to capture \emph{malleable} VNRs, where each request can be instantiated using one of several functionally equivalent topologies that trade resources differently. This flexibility can improve embedding feasibility, but only if the orchestrator can jointly select suitable alternatives and embed them under dynamic arrivals. This paper proposes \textbf{HRL-VNEAP}, a hierarchical reinforcement learning approach for dynamic VNEAP. A high-level policy selects the most suitable alternative topology (or rejects the request), and a low-level policy embeds the chosen topology onto the substrate network. Experiments on realistic substrate topologies under varying arrival rates show that naive exploitation strategies provide only modest gains, whereas HRL-VNEAP outperforms state of the art approaches, improving acceptance ratio by up to 22\%, and net profit by up to 20\%. An
offline MILP upper bound is also used on tractable instances to quantify the remaining optimality gap.
\end{abstract}


\titlespacing*{\section}
  {0pt}{*0.8}{*0.4} 

\section{Introduction}
Network slicing enables the partition of a shared physical network into multiple, logically isolated \lq \lq slices'' each tailored to a tenant or service class with distinct requirements, for instance, in terms of bandwidth, reliability, and security \cite{networkslicing}.

A central problem in network slicing is the \emph{Virtual Network Embedding (VNE)} problem \cite{fischer}, which concerns the mapping of Virtual Network Requests (VNR) onto a Substrate Network (SN). In this context, each slice request, i.e., each VNR, specifies the virtual nodes and connections it requires, along with their associated resource demands. The objective of the VNE is to allocate resources of the VNR (i.e., to embed the VNR) onto physical nodes and links while ensuring resource isolation and adherence to quality-of-service requirements (e.g., minimum bandwidth guarantees, maximum latency constraints, or packet loss limits). The embedding must also optimize system-level goals such as cost efficiency and resource utilization.

\begin{figure}[t]  
        \centering
        \includegraphics[width=0.9\linewidth]{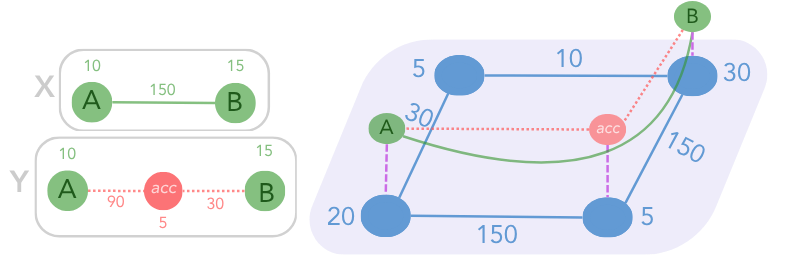}

\caption{\footnotesize VNR instance with two functionally-equivalent alternatives. X realizes the service as a direct two-node VN (A: 10, B: 15 CPU; 150 BW). Y inserts an intermediate accumulator (5 CPU) that compresses traffic, reducing aggregate bandwidth from 150 to 120 (90+30) at the cost of additional compute. The right panel shows one possible substrate embedding of X and Y.}

    \label{fig:alt1}
\end{figure}

Standard VNE models treat each VNR as a single virtual graph whose nodes,
links, and resource demands are fixed before embedding~\cite{fischer}. As a
result, the embedding algorithm can only decide where to place this given graph
on the SN; it cannot choose a different virtual topology for the same service,
even if that topology would be easier to embed under the current resource
availability. To overcome this rigidity, recent work has introduced \emph{malleable} VNRs~\cite{malleable}, where each request can be realized through a set of several functionally equivalent virtual topologies (\emph{alternatives}), that trade computation and bandwidth differently. Fig.~\ref{fig:alt1} illustrates this principle on a minimal example: the same service can be realized either as a direct two-node topology (X)
or with an intermediate \emph{accumulator} node (Y) that reduces bandwidth at the cost of additional CPU. This flexibility creates a significant practical advantage as it allows the embedding algorithm to select a topology aligned with the current state of the SN, thereby improving overall network efficiency \cite{bari}.

Compared to classical VNE, the VNE with Alternatives Problem (VNEAP) jointly optimizes the selection of an alternative and its embedding. This adds a new decision layer: the orchestrator has to select which topology to instantiate and then to embed it onto the SN. Consequently, the benefit of alternatives depends on the decision mechanism. Naive selection rules, such as choosing highest revenue alternatives, may saturate the network early and make future embeddings difficult. Therefore, alternatives do not automatically improve network performance; they become useful \emph{only} when the orchestration mechanism exploits their resource diversity by matching each alternative to the current SN state while accounting for long-term resource availability. Reinforcement Learning (RL) is well suited to this dynamic setting because VNRs arrive online and each embedding decision affects future substrate resources, making the problem sequential. However, a flat policy must learn alternative selection and embedding jointly, although these decisions have different action spaces and time scales. This motivates our adopted hierarchical design, where alternative selection and embedding are optimized jointly by separate yet coupled RL policies.


\begin{figure}[t]
    \centering
    \includegraphics[width=0.8\linewidth]{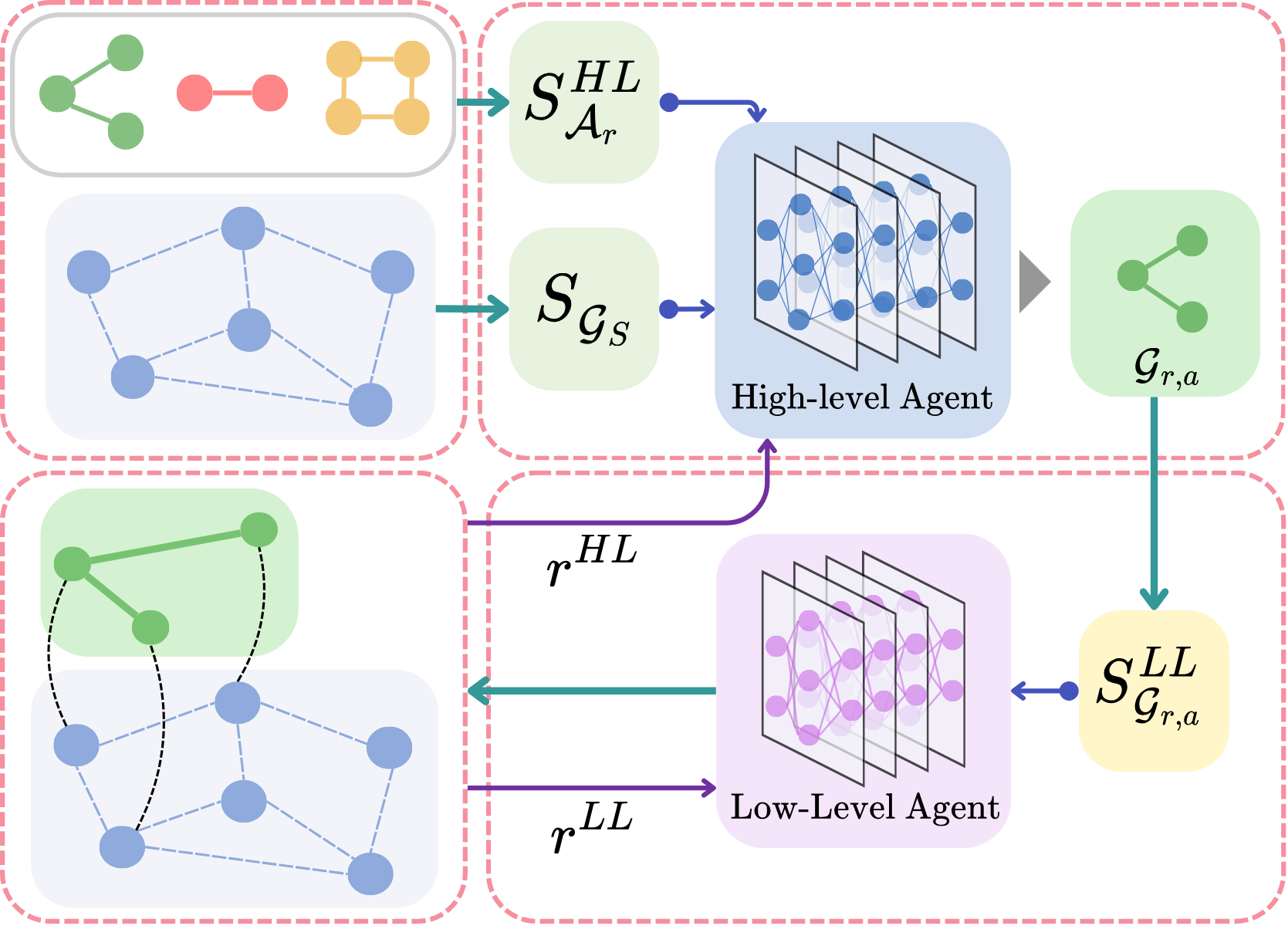}
    \caption{\footnotesize Schematic representation of the proposed HRL architecture. The HL agent observes the substrate and request features to select the optimal alternative. Subsequently, the LL agent performs the embedding. Distinct reward signals ($r^{HL}$ and $r^{LL}$) are used to update the respective policies.}
    \label{fig:arch}
\end{figure}

In this work, we focus on VNEAP in the dynamic regime where VNRs arrive online and the orchestrator must act without knowledge of future requests. We propose \emph{HRL-VNEAP}, a hierarchical RL framework that decomposes the decision process into two levels. A high-level (HL) policy selects an alternative or rejects the request and a low-level (LL) policy embeds the selected alternative through sequential node placement and link routing. This decomposition allows each policy to operate over a more focused decision space: the HL agent learns alternative-selection behavior under substrate dynamics, while the LL agent specializes in embedding quality under substrate constraints. 

Because dynamic VNEAP lacks an established online optimal benchmark, we also formulate an offline Mixed Integer Linear Programming (MILP) with full knowledge of the request sequence. This MILP is not an achievable online target; rather, it provides an offline upper bound which we use to quantify the gap between online learning and the optimal solution on tractable instances. Moreover, we instantiate the proposed framework through a Musical Metaverse (MM) use case. MM is a representative VNEAP scenario because the same interactive networked musical service can be realized through different media-processing configurations, each trading CPU, bandwidth, and delay differently \cite{turchet2023musical}. Our contributions are summarized as follows:

\begin{itemize}
    \item We study VNEAP under online VNR arrivals where alternative selection and embedding decisions must be made
    without future knowledge.

    \item We propose HRL-VNEAP, a two-level RL framework that explicitly decomposes the problem into
    alternative selection and VNR embedding. 

    \item We develop an offline MILP formulation for VNEAP and use it as a theoretical upper bound to quantify the remaining gap of online learning-based decisions.

    \item Since the generation of VNR alternatives is left outside the scope of prior VNEAP work~\cite{malleable}, we instantiate realistic alternatives through a MM use case, deriving functionally equivalent topologies from real-time multiparty media architectures \cite{rottondi}.
\end{itemize}

\section{Related Work}
\label{sec:related}

The VNE problem has been widely studied through exact methods, heuristics, and learning-based approaches. Surveys such as Fischer \emph{et al.}~\cite{fischer} and Wu \emph{et al.}~\cite{10587211} summarize these line of works, while recent work has focused in particular on RL-based models~\cite{lim} to overcome the intractability of exact VNE formulations.

HRL has also been explored in VNE to improve scalability by splitting the decision process. In VNE-HRL~\cite{vnehrl}, a high-level agent selects which VNR to embed next, while a low-level agent performs the actual mapping. HRL-ACRA~\cite{wang2023joint} extends this idea to joint admission control and embedding, again using a high-level policy for admission decisions and a low-level policy for resource allocation. Both works show that hierarchical decomposition improves long-term reward and resource utilization over flat RL methods, but both target classical VNE with predetermined VNR topologies. The high-level decision in VNEAP is structurally different: rather than ordering requests or admitting a fixed VNR, the agent must choose among functionally equivalent topologies whose resource profiles differ and reason jointly over substrate state and per-alternative trade-offs.

Motivated by emerging applications that allow multiple service or
function configurations, VNEAP~\cite{malleable} introduces configurable
alternatives for each VNR. The original work focuses on handcrafted heuristics
for static settings and assumes that the alternative set is predefined, while
no learning-based method has addressed the dynamic VNEAP decision space. Our
work fills this gap by introducing an HRL solution that learns to exploit alternatives where naive or fixed selection rules fail, and by
benchmarking it against an offline MILP upper bound with full knowledge of the
request sequence.

\begin{table}[tbp]
\caption{\footnotesize System model sets, parameters and variables}
\label{tab:sets}
\centering
\scriptsize
\renewcommand{\arraystretch}{0.5}%
\begin{tabular}{|lp{5cm}|}
\hline
\emph{Sets and Notations} & \emph{Description} \\ 
\hline
$\mathcal{G}_s=(\mathcal{N}_s,\mathcal{E}_s)$ & Directed substrate network graph  \\
$\mathcal{R} = \{r_1, ...,r_n\} $ & Set of virtual network requests \\
$\mathcal{A}_{r} = \{\mathcal{G}_{r,1}, ...,\mathcal{G}_{r,|\mathcal{A}_r|}\} $ & Set of alternatives of VNR $r \in \mathcal{R}$\\
$\mathcal{G}_{r,a} = (\mathcal{N}_{r,a},\mathcal{E}_{r,a})$ & Virtual topology of alternative $a \in \mathcal{A}_r$ \\
$T = \{1, ..., |T|\}$ & Set of time events \\
$\mathcal{S}_{r,a,i}$ & Set of $\mathcal{N}_s$ allowed to host virtual node $i \in \mathcal{N}_{r,a}$\\
$\rho_{pq}$ & Set of substrate paths from $p$ to $q$ \\
$P$ & A path $\in \rho_{pq}$ \\

\hline
\emph{Parameters} & \emph{Description} \\ 
\hline

$c_{p} \in \mathbb{R}^{+}$ & Capacity of node $p \in \mathcal{N}_s$ \\
$b_{uv} \in \mathbb{R}^{+}$ & Bandwidth of link $(u,v) \in \mathcal{E}_s$ \\
$l_{uv} \in \mathbb{R}^{+}$ & Propagation delay of link $(u,v) \in \mathcal{E}_s$ \\
$c_{r,a,i} \in \mathbb{R}^{+}$ & Capacity demand of node $i \in \mathcal{N}_{r,a}$ \\
$b_{r,a,ij} \in \mathbb{R}^{+}$ & Bandwidth of link $(i,j) \in \mathcal{E}_{r,a}$ \\
$d_{r,a,ij} \in \mathbb{R}^{+}$ & Maximum tolerable delay of link $(i,j) \in \mathcal{E}_{r,a}$ \\
$\tau_{r} \in T$ & Arrival time of request $r \in \mathcal{R}$ \\
$\lambda_r \in \mathbb{R}^{+}$ & Lifetime (holding) of request $r \in \mathcal{R}$ \\ 
$\delta(P) \in \mathbb{R}^{+}$ & Propagation delay of substrate path P \\
$\alpha_{r,t} \in \{0,1\}$ & Active flag variable \\

\hline
\emph{Variables} & \emph{Description} \\
\hline

$y_{r,a} \in \{0,1\}$ & Selection of alternative variable \\
$x^{p}_{r,a,i} \in \{0,1\}$ & Node placement variable\\
$w^{P}_{r,a,ij} \in \{0,1\}$ & Link routing variable\\
\hline
\end{tabular}
\end{table}

\section{Dynamic VNEAP}
We consider a SN modeled as a graph $\mathcal{G}_s=(\mathcal{N}_s,\mathcal{E}_s)$, where each substrate
node $p \in \mathcal{N}_s$ has computing capacity $c_p$, and each substrate link $(u,v) \in \mathcal{E}_s$ has bandwidth capacity
$b_{uv}$ and a processing delay $l_{uv}$.
A set of VNRs $\mathcal{R}$ arrives over time. Each request $r \in \mathcal{R}$ is characterized by an arrival time $\tau_r$
and a holding time (lifetime) $\lambda_r$. Unlike \emph{standard} VNE, each request carries a set of \emph{functionally equivalent}
alternatives
$\mathcal{A}_r = \{\mathcal{G}_{r,1}, \mathcal{G}_{r,2}, \dots,
\mathcal{G}_{r,|\mathcal{A}_r|}\}$, where each alternative
$\mathcal{G}_{r,a} = (\mathcal{N}_{r,a}, \mathcal{E}_{r,a})$ is a
directed graph of virtual nodes and links. Every virtual node $i \in \mathcal{N}_{r,a}$ represents a VNF with CPU demand $c_{r,a,i}$, and every virtual link
$(i,j) \in \mathcal{E}_{r,a}$ represents inter-VNF communication with bandwidth demand $b_{r,a,ij}$, and a maximum tolerable propagation delay $d_{r,a,ij}$.\\
The VNEAP orchestrator jointly decides, for each $r \in \mathcal{R}$, whether to admit it and which alternative $\mathcal{G}_{r,a}$ to instantiate, together with the embedding of the chosen alternative onto the substrate $\mathcal{G}_s$, subject to capacity, routing, and delay constraints. The optimization objective is to maximize net profit (NP) defined in Eq.~(\ref{eq:a}). Tab.~\ref{tab:sets} summarizes the notation used in the formulation. The MILP is described as follows:

{\small
\setlength{\jot}{2pt}
\begin{align}
\max\ & \sum_{r\in\mathcal{R}}\sum_{a\in\mathcal{A}_r}\!\big(y_{r,a}\,\mathrm{Rev}_{r,a} - \mathrm{Cost}_{r,a}\big)  \label{eq:a} \tag{a}\\
\text{s.t.}\ & \sum_{a\in\mathcal{A}_r} y_{r,a} \leq 1 \qquad \forall r \label{eq:b} \tag{b}\\
& \sum_{p\in\mathcal{S}_{r,a,i}} x^{p}_{r,a,i} = y_{r,a} \qquad \forall r,a,\,i\in\mathcal{N}_{r,a} \label{eq:c} \tag{c}\\
& \sum_{i\in\mathcal{N}_{r,a}} x^{p}_{r,a,i} \leq y_{r,a} \qquad \forall r,a,\,p\in\mathcal{N}_s \label{eq:d} \tag{d}\\
& \sum_{\substack{p,q\in\mathcal{N}_s\\ q\neq p,\,P\in\rho_{pq}}}\!\! w^{P}_{r,a,ij} = y_{r,a} \quad \forall r,a,(i,j)\in\mathcal{E}_{r,a} \label{eq:e} \tag{e}\\
& \sum_{\substack{q\in\mathcal{N}_s\setminus\{p\}\\ P\in\rho_{pq}}}\!\! w^{P}_{r,a,ij} \leq x^{p}_{r,a,i} \quad \forall r,a,(i,j),p \label{eq:f} \tag{f}\\
& \sum_{\substack{p\in\mathcal{N}_s\setminus\{q\}\\ P\in\rho_{pq}}}\!\! w^{P}_{r,a,ij} \leq x^{q}_{r,a,j} \quad \forall r,a,(i,j),q \label{eq:g} \tag{g}\\
& \sum_{r,a,i} \alpha_{r,t}\,c_{r,a,i}\,x^{p}_{r,a,i} \leq c_{p} \qquad \forall p\in\mathcal{N}_s,\,t\in T \label{eq:h} \tag{h}\\
& \sum_{r,a,(i,j)}\!\!\alpha_{r,t}\,b_{r,a,ij}\!\!\sum_{\substack{p\neq q\\ P\in\rho_{pq},\,(u,v)\in P}}\!\!\! w^{P}_{r,a,ij} \leq b_{uv} \quad \forall (u,v)\in\mathcal{E}_s \label{eq:i} \tag{i}\\ 
& \sum_{\substack{p,q\in\mathcal{N}_s,\,q\neq p\\ P\in\rho_{pq}}}\!\! w^{P}_{r,a,ij}\,\delta(P) \leq d_{r,a,ij} \quad \forall r,a,(i,j) \label{eq:j} \tag{j}\\
& y_{r,a},\,x^{p}_{r,a,i},\,w^{P}_{r,a,ij} \in \{0,1\} \label{eq:k} \tag{k}
\end{align}
}

$\text{Rev}_{r,a}$ represents the economic value of the allocated virtual resources and defined as:
\begin{small}
\begin{equation*}
\text{Rev}_{r,a} = \lambda_r \cdot \left(
\pi_c \sum_{i\in\mathcal{N}_{r,a}} c_{r,a,i}
+ \pi_b \sum_{(i,j)\in\mathcal{E}_{r,a}} b_{r,a,ij}
\right)
\end{equation*}
\end{small}

where $\pi_c$ and $\pi_b$ denote the tenant price per CPU and bandwidth unit, respectively. $\text{Cost}_{r,a}$ captures the actual consumption of physical resources.
\begin{small}
\begin{equation*}
\begin{aligned}
\text{Cost}_{r,a} = \lambda_r \cdot \bigg(
& \kappa_c \sum_{i\in\mathcal{N}_{r,a}} \sum_{p\in\mathcal{N}_s}
x^p_{r,a,i}\, c_{r,a,i} \\
& + \kappa_b \sum_{(i,j)\in\mathcal{E}_{r,a}}
\sum_{\substack{p,q\in\mathcal{N}_s\\q\neq p}}
\sum_{P\in\mathcal{P}_{pq}}
w^P_{r,a,ij}\, b_{r,a,ij}\, |P|
\bigg)
\end{aligned}
\end{equation*}
\end{small}
where $\kappa_c$ and $\kappa_b$ denote the operator cost per CPU and bandwidth unit.

Eq.~(\ref{eq:b}) ensures that, for each VNR $r \in \mathcal{R}$, at most one alternative is selected. Eq.~(\ref{eq:c}) ensures that each virtual node $i \in \mathcal{N}_{r,a}$ of a selected alternative $\mathcal{G}_{r,a}$ (i.e., when $y_{r,a} =1$), is mapped on exactly one substrate node $p \in \mathcal{S}_{r,a,i}$. Eq.~(\ref{eq:d}) ensures that no two virtual nodes of the same VNR are co-located over the same substrate node. Eq.~(\ref{eq:e}) is the link routing constraint, across all candidate endpoint pairs $(p,q)$ and their candidate paths, exactly one path for each virtual $(i,j) \in \mathcal{E}_{r,a}$ is selected, for the selected topology.
Eq.~(\ref{eq:f}) and Eq.~(\ref{eq:g}) are the source and destination endpoint coupling constraints (path must start where $i$ is placed, and must end where $j$ is placed, respectively). We define an activity flag $\alpha_{r,t}$ to indicate whether request $r$ is active at time $t$, so that only active VNRs consume SN resources in the capacity constraints:
\begin{equation}
    \alpha_{r,t} = \begin{cases}
    1 \quad \text{if} \quad t \in \{\tau_r, \tau_r+1, \dots, \tau_r + \lambda_r -1 \} \\ 
    0 \quad \text{otherwise}
\end{cases}
\label{eq:d1}
\end{equation}
Eq.~(\ref{eq:h}) ensures that at each time slot $t \in T$, the total active capacity demand placed on $p \in \mathcal{N}_s$ cannot exceed its maximum capacity $c_p$. Eq.~(\ref{eq:i}) ensures link bandwidth constraint, where for each substrate link $(u,v) \in \mathcal{E}_S$ and time $t$, sum of the bandwidth demands of all active virtual links that chose a path containing $(u,v)$, must not exceed the substrate link bandwidth $b_{uv}$. Eq.~(\ref{eq:j}) ensures that the substrate path allocated to
each virtual link does not exceed its delay tolerance, where
$\delta(P)$ denotes the propagation delay of substrate path~$P$. \\
\emph{Complexity:} The standard VNE problem is known to be $\mathcal{NP}$-hard \cite{fischer}, and the introduction of alternatives further expands the search space. Solving this MILP using exact solvers becomes computationally prohibitive for large problem instances.

\section{Hierarchical RL for VNEAP}
\label{sec:hrl} 

HRL-VNEAP decomposes dynamic VNEAP into two sequential decisions over a shared view of SN (Fig.~\ref{fig:arch}). For each arriving VNR, an HL agent selects an alternative from $\mathcal{A}_r$ or rejects the request; if selected, a LL agent embeds the chosen alternative through sequential node placement and link routing. Both agents share the substrate observation block, so the HL agent's selection is informed by the same embeddability features that the LL agent operates on, reducing the cross-level coordination challenges noted in the HRL literature~\cite{nachum2018data}. Each level is updated by a tailored reward, $r^{HL}$ and $r^{LL}$, allowing specialized credit assignment despite a coupled decision process. The rest of this section details state, action, and reward design at each level, followed by policy and training details.
\begin{table}[t]
    \centering
    \scriptsize
    \setlength{\tabcolsep}{2.5pt}
    \renewcommand{\arraystretch}{0.82}
    \caption{Observation features of the high-level and low-level policies.}
    \label{tab:states}
    \begin{tabularx}{\columnwidth}{@{}llX@{}}
        \toprule
        \textbf{Block} & \textbf{Policy} & \textbf{Features} \\
        \midrule
        Substrate 
        & Both 
        & \textit{Nodes:} rem. cap., load, betweenness centrality degree; 
          \textit{Links:} rem. BW, load, betw. cent.; \textit{node/link indices for masking.} \\
        
        Virtual 
        & HL 
        & \#nodes, \#links, avg. cap., avg. BW, avg. path length, avg. connectivity per alternative, min. and max. tolerable latencies. \\
        
        Virtual 
        & LL 
        & Node cap. demands, link-BW sums, node degrees, tolerable latencies. \\
        
        Context 
        & HL 
        & VNR lifetime. \\
        
        Context 
        & LL 
        & Progress ratio, current-node cap., current-node degree. \\
        \bottomrule
    \end{tabularx}
\end{table}

\subsection{High-level agent}
The \emph{high-level state space $S^{HL} = S^{HL}_{\mathcal{A}_r} || S_{\mathcal{G}_s}$} is reported in Tab.~\ref{tab:states}, and consists of encoded information from both $\mathcal{G}_s$ and $\mathcal{A}_r$, including the topological description of each alternative, requested resources, and the lifetime of each VNR. 

The \emph{high-level action space} $A^{HL}=\{1,2, \dots,|\mathcal{A}_r|+1 \}$ is designed as a discrete variable that indicates the selected alternative, in addition to the action of rejecting a VNR (meaning that none of its alternatives is selected). We denote the sampled action at time $t$ as $a^{HL}_t$.

The \emph{High-level reward $R^{HL}$} is designed to guide the agent to optimize the long-term resource efficiency. We define $r_{t}^{HL} \in R^{HL}$ as follows:\noindent
\begin{equation}
r_{t}^{HL} = \begin{cases}
    \text{Rev}(a^{HL}_t) - \text{Cost}(a^{HL}_t) \quad \text{if $a^{HL}_t$ is embedded} \\
    -\text{Rev}(a^{HL}_t) \quad \text{if $a^{HL}_t$ is admitted but not embedded} \\
    0 \quad\text{otherwise}
\end{cases}
\label{eq:rhl}
\end{equation}

The HL reward in Eq.~(\ref{eq:rhl}) mirrors the per-request net profit optimized by the MILP (Eq.~\ref{eq:a}), aligning the learning signal with the objective. Successful embeddings are rewarded by their net profit, encouraging selection of profitable alternatives. In case of failed embedding, the agent receives a negative reward. In case the agent's rejects a VNR, it is neither rewarded nor penalized.


\subsection{Low-level agent}
The \emph{Low-level state space $S^{LL} = S^{LL}_{\mathcal{G}_{r,a}} || S_{\mathcal{G}_s}$} shares the same encoded SN information with the HL state, thus both agents operate on an identical view of the SN. Since the LL agent's objective is to perform the embedding of the chosen topology per VNR, it receives specific information of the requested resources (see Tab.~\ref{tab:states}). To reflect the sequential nature of the embedding procedure, the state also includes specific context variables that indicate which component of the VNR is currently being embedded such as node index.

The \emph{Low-level action space $A^{LL}$} is defined as $A^{LL}=\{1,2, \dots,|\mathcal{N}_s| \}$ where each element corresponds to a specific node within the SN. At each step $t$, the agent will select an action $a^{LL}_{t,k}$ that corresponds to a substrate node where node $k$ of alternative $a^{HL}_t$ will be embedded. The allocation of links is done progressively via K-Shortest Path (K-SP) \cite{AutomaticHRL, ksp}.

For the \emph{Low-level reward $R^{LL}$} we adopt the multi-objective reward of \cite{AutomaticHRL}, which decomposes into placement-cost and load-balancing components to alleviate sparse-reward exploration.





\subsection{Policy Network and Training Details}

Both agents are trained using Proximal Policy Optimization (PPO)~\cite{schulman2017proximal}. Because the VNEAP observation combines substrate-state features, alternative-level VNR features, and embedding-step context, we replace the default PPO feature extractor with a compact Multi Layer Perceptron (MLP) based encoder\footnote{All implementation details will be released with the source code in case of paper acceptance.}. Each feature group is encoded by a dedicated branch and projected to a common latent space. The resulting representation is passed to separate actor and critic heads. Tab.~\ref{tab:hyperparams} summarizes the main training and architecture hyperparameters.



\begin{table}[t]
\centering
\scriptsize
\caption{PPO training and policy-network hyperparameters.}
\label{tab:hyperparams}
\setlength{\tabcolsep}{3pt}
\renewcommand{\arraystretch}{0.88}
\begin{tabular}{@{}ll@{}}
\toprule
\textbf{Hyperparameter} & \textbf{Value} \\
\midrule
Policy extractor & MLP branches for substrate, VNR, context \\
Hidden layers & $[256,256,256]$ \\
Activation & ReLU \\
Learning rate & $3\times10^{-4}$ \\
Discount factor $\gamma$ & $0.99$ \\
Entropy coefficient & $0.01$ \\
Batch size & 128 \\
Rollout steps & 10 \\
Optimizer & AdamW \\
\bottomrule
\end{tabular}
\end{table}

\begin{table}[t]
\centering
\scriptsize
\caption{Sampling distributions for substrate, request, and
per-stream resource coefficients.}
\label{tab:sampling}
\setlength{\tabcolsep}{4pt}
\renewcommand{\arraystretch}{0.92}
\begin{tabular}{@{}ll@{\hspace{2em}}ll@{}}
\toprule
\textbf{Parameter} & \textbf{Range} & \textbf{Parameter} & \textbf{Range} \\
\midrule
Substrate CPU/link cap.  & $U[50,100]$    & Mixing CPU $\mu_r$       & $U[4,8]$    \\
Stream bandwidth $b_r$   & $U[4,10]$       & Forwarding CPU $\phi_r$  & $U[1,3]$\\
Endpoint CPU $e_r$       & $U[1,6]$       & Link delay $\ell_{uv}$   & $U[1,5]$ ms \\
Lifetime $\lambda_r$     & Exp., mean $50$& Sync.\ budget $D_r^{\max}$ & $30$ ms   \\
\bottomrule
\end{tabular}
\end{table}

\section{Experimental Setup}
We evaluate HRL-VNEAP on two substrate topologies, Atlanta and GEANT~\cite{SNDlib10}. Simulation parameters are reported in Tab.~\ref{tab:sampling}. Each simulation contains $50$ dynamically arriving VNRs following a Poisson process of rate $\eta$.  

\subsection{Musical Metaverse alternatives}
\begin{figure}[t]
    \centering
    \includegraphics[width=0.94\linewidth]{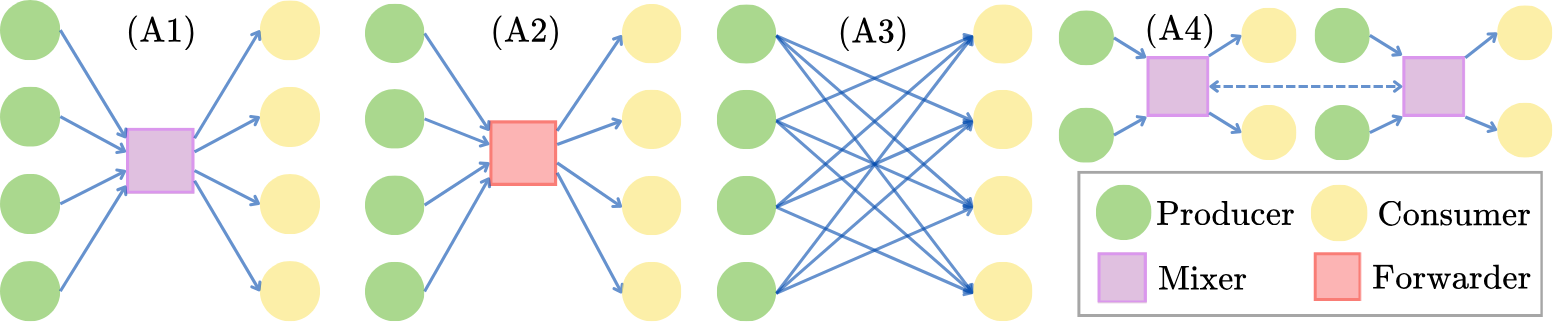}
    \caption{MM alternatives: (A1) centralized immersive mixer, (A2) SFU-style forwarder, (A3) peer-to-peer full mesh, and (A4) clustered immersive mixer. Green nodes denote musician's producer endpoints (generates data),
    yellow nodes denote musician's consumer endpoints (renders data), and squares denote media VNFs.
    In (A3), each endpoint acts as both producer and consumer.}
    \label{fig:alternatives}
\end{figure}

To instantiate VNR alternatives, we use a MM scenario in which each VNR represents a synchronized musical-performance slice with $n_r =4$ musicians. Each VNR has four functionally equivalent alternatives, as shown in Fig.~\ref{fig:alternatives}, which preserve the same service semantics where every musician receives the contribution of the others within a synchronization budget $D_r^{\max}$ but differ in where specific functions, such as audio processing and stream distribution, are placed ~\cite{tsioutas2022assessing,boem2025audio}. 
\textbf{A1 (Centralized mixer):} A single mixing VNF receives one stream of bandwidth $b_r$ from each musician, mixes the $n_r$ streams with CPU demand $n_r\mu_r$, and returns one mixed stream of bandwidth $b_r$ to each musician \cite{rottondi, turchet2023musical}.\\
\textbf{A2 (Selective Forwarding Unit):} A forwarding VNF receives one stream $b_r$ from each musician and forwards streams without mixing. Its CPU demand is $n_r\phi_r$, with $\phi_r<\mu_r$. Each egress carries $(n_r-1)b_r$, and endpoint mixing requires CPU $e_r+(n_r-1)\mu_r$ \cite{stickland2019design}.\\
\textbf{A3 (Peer-to-peer mesh):} No central VNF is used. Each musician sends one stream $b_r$ directly to every other musician over directed peer-to-peer links and performs local mixing with CPU demand $e_r+(n_r-1)\mu_r$ \cite{caceres2010jacktrip}.\\
\textbf{A4 (Clustered mixer):} Musicians are partitioned into two clusters, each served by a local mixing VNF with CPU demand $(\frac{n_r}{2}+1)\mu_r$. Local links carry bandwidth $b_r$, and the two mixers exchange aggregate partial mixes over bidirectional inter-cluster links of bandwidth $b_r$ \cite{rottondi}.
These four alternatives cover complementary CPU/bandwidth trade-offs: A1 concentrates compute centrally, A2 trades compute for replicated egress bandwidth, A3 removes central infrastructure at the cost of $O(n_r^2)$ links, and A4 distributes compute across local mixers while adding an inter-cluster synchronization path. To match the per-virtual-link delay constraint in Eq.~(\ref{eq:j}), we assign delay tolerances according to each template's critical path: $15$ ms for A1 and A2, $30$ ms for A3, and $10$ ms for A4. 

\subsection{Performance under increasing arrival rate}
\label{subsec:baselines}

Our evaluation tests the extent to which HRL-VNEAP can convert alternatives into performance gains. We first compare against established state of the art (SOTA) VNE approaches under dynamic arrivals. Since these baselines cannot choose among
multiple VNR alternatives by themselves, we pair them with simple selectors that choose one alternative before embedding. We also perform an analysis by varying the number of alternatives per VNR to assess whether additional malleability is actually exploited. Finally, we report runtime to evaluate the computational cost of exploiting this flexibility.

We compare HRL-VNEAP with four VNE algorithms: NEA-RANK~\cite{nea} (NEA), PG-CNN~\cite{pg-cnn} (PG), HRL-ACRA \cite{wang2023joint} (ACRA), and Flag-VNE \cite{wang2024flagvne} (Flag). Since these methods assume one topology per request, we adapt them to VNEAP through a two-stage wrapper where a selector first chooses one alternative $a\in\mathcal{A}_r$, and the chosen alternative is then embedded using the corresponding VNE algorithm. We consider two selectors: \emph{Random}, which samples an alternative randomly from $\mathcal{A}_r$, and \emph{Max-Revenue}, which selects
$a^\star = \arg\max_{a\in\mathcal{A}_r} \mathrm{Rev}_{r,a}$.
Random represents availability without informed exploitation, while Max-Revenue represents a greedy rule that ignores embedding difficulty and future resource impact. We also report variants using our LL embedding policy with these selectors to isolate the benefit of hierarchical training. In the figures, ``-r'' and ``-m'' denote Random and Max-Revenue selection,
respectively. All reported results are averaged over $50$ independent runs with different random seeds.
\begin{figure}[t]

    \includegraphics[width=0.95\linewidth]{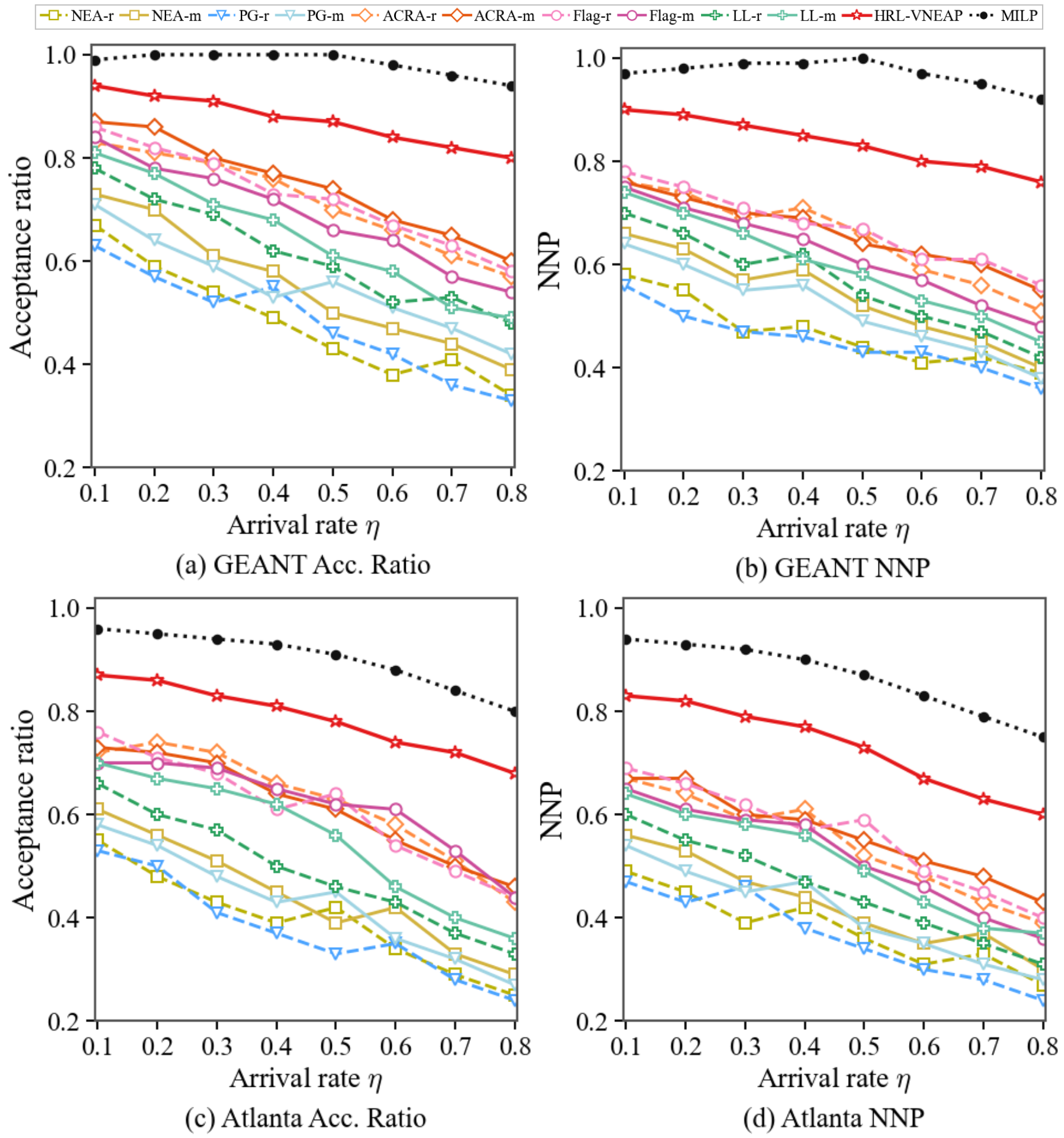}
    \caption{Performance under increasing traffic load on GEANT and Atlanta:
    acceptance ratio and normalized net profit (NNP) versus arrival rate
    $\eta$.}
    \label{fig:part1}
\end{figure}

\section{Results and Discussion}
\subsection{Performance Gains of Coordinated HRL}
Fig.~\ref{fig:part1} reports acceptance ratio and normalized net profit (NNP) Eq.~(\ref{eq:a})
for arrival rates $\eta\in[0.1,0.8]$ on GEANT and Atlanta SN topologies \cite{SNDlib10}. The results show that HRL-VNEAP consistently outperforms all SOTA approaches across both SNs and in terms of both acceptance ratio and NNP. This confirms that the proposed design is well suited to the VNEAP setting, where the algorithm can jointly account for embedding feasibility, resource efficiency, and the availability of alternatives. Results also show that the advantage of HRL-VNEAP becomes more evident under higher arrival rate. As the arrival rate increases, all methods degrade because the SN resources become increasingly constrained. However, HRL-VNEAP maintains relatively high performance by exploiting the available alternatives more effectively. For instance, at the highest arrival rate, HRL-VNEAP still achieves around 0.80 acceptance on GEANT and 0.68 on Atlanta, while the strongest SOTA remain clearly lower with 0.62 and 0.51 respectively.\\
The results on Atlanta further confirm the robustness of our proposed approach. Competing methods exhibit a sharper performance decline, whereas HRL-VNEAP maintains a clear margin over the baselines, demonstrating that the learned policy remains effective even when fewer embedding options are available. The offline MILP upper bound remains above all online methods because it has full future knowledge. HRL-VNEAP is nevertheless the closest online method, indicating that the hierarchical policy captures much of the value of alternative topologies while operating without future arrivals. 

\subsection{Impact of the Number of Alternatives}

\begin{figure}
    \includegraphics[width=0.9\linewidth]{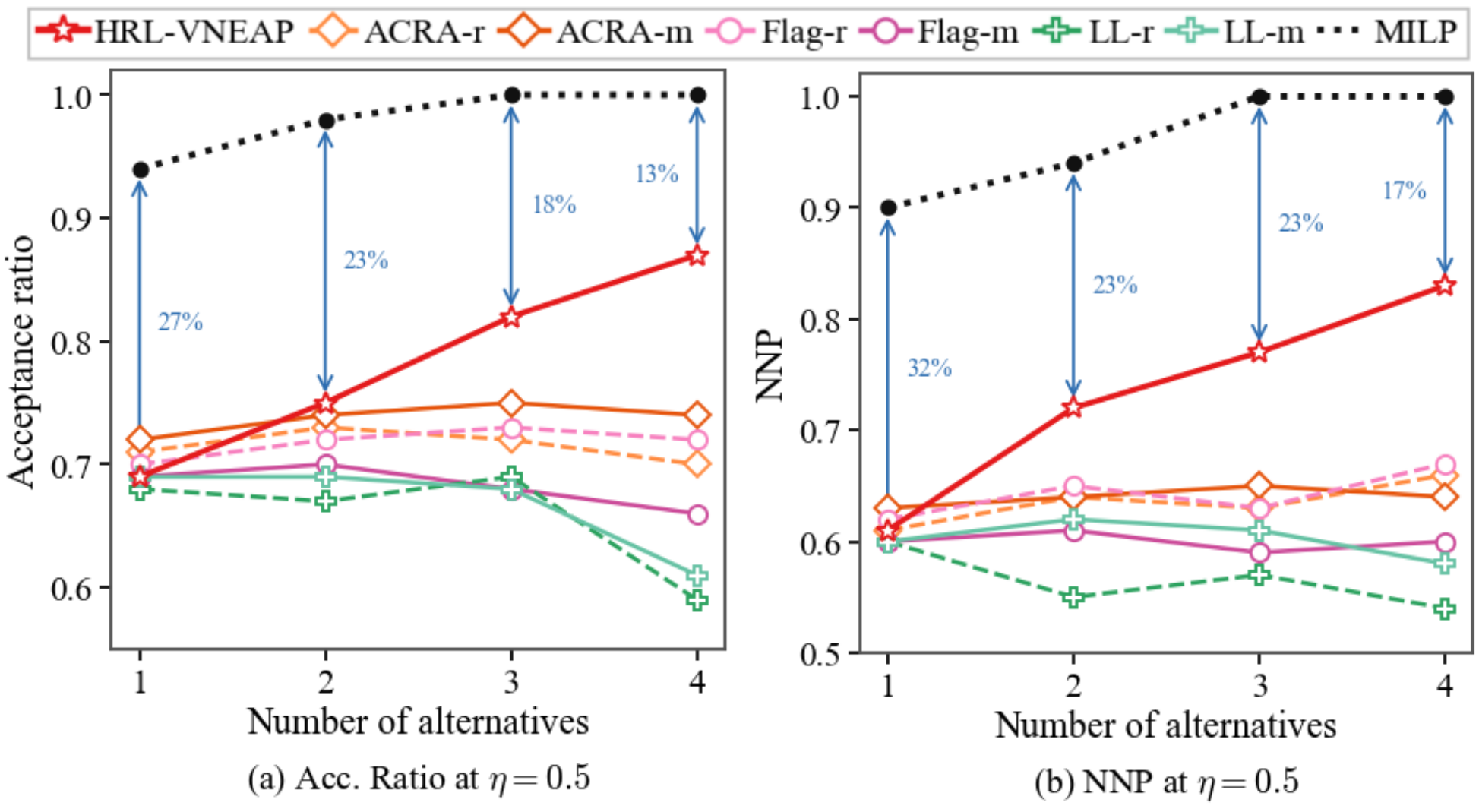}
    \caption{Impact of the number of alternatives on GEANT at $\eta=0.5$. Alternatives are selected randomly.}
    \label{fig:part2}
\end{figure}

Fig.~\ref{fig:part2} reports the impact of the number of alternatives on acceptance ratio and NNP at $\eta=0.5$. Due to space limitations, we report results on GEANT. As can be expected, having more alternatives allows improving the acceptance ratio, since the embedding algorithm has more candidate requests from which to select a feasible and resource-compatible option. However, the gain is not uniform across the various approaches, showing that alternatives can be useful only when they are effectively exploited by a robust policy. 
With one alternative, the performance of
HRL-VNEAP is comparable to the strongest baselines, as expected, since the
alternative-selection dimension is almost absent and the problem approaches VNE. As $|\mathcal{A}_r|$ increases, HRL-VNEAP
benefits much more clearly from the additional flexibility. With four
alternatives, it reaches about $0.87$ acceptance ratio and $0.83$ NNP,
compared with roughly $0.74$ and $0.67$ for the strongest baseline,
corresponding to gains of approximately $17.6\%$ and $23.9\%$, respectively.
The random-selection variants fluctuate because the selected topology may not
match the current SN state, while Max-Revenue variants saturate since
they favor immediate reward without accounting for embedding difficulty or
future resource impact. The MILP upper bound also improves with
$|\mathcal{A}_r|$, confirming that additional alternatives enlarge the
attainable solution space. HRL-VNEAP follows this trend most closely among
online methods: its gap to the MILP decreases from about $27\%$ to $13\%$ in
acceptance ratio and from about $31\%$ to $17\%$ in NNP as the number of
alternatives increases from one to four.



\subsection{Exploitation of alternatives and Runtime}
In Fig.~6(a) each MM alternative is evaluated alone on GEANT at $\eta=0.5$. The results show that selecting only one fixed alternative leads to very similar acceptance ratios (0.71 - 0.75). This indicates that no single template is consistently sufficient to unlock a large gain by itself. In contrast, when all alternatives are jointly available, the acceptance ratio increases to $0.87$. This confirms that the benefit does not come only from defining multiple VNRs, but from allowing the embedding algorithm to exploit their complementarity and select the most suitable realization according to the current network state.

Fig.~6(b) reports the runtime comparison on GEANT and Atlanta. As expected, MILP is computationally expensive, reaching seconds of runtime, especially on the larger GEANT topology. In contrast, the learning-based methods return decisions in the order of milliseconds. HRL-VNEAP introduces only a negligible runtime increase compared with the fastest baselines. This small additional cost is justified by the performance gains reported earlier, where HRL-VNEAP achieves higher acceptance and NNP by learning how to exploit alternatives more effectively.

\begin{figure}
    \includegraphics[width=0.9\linewidth]{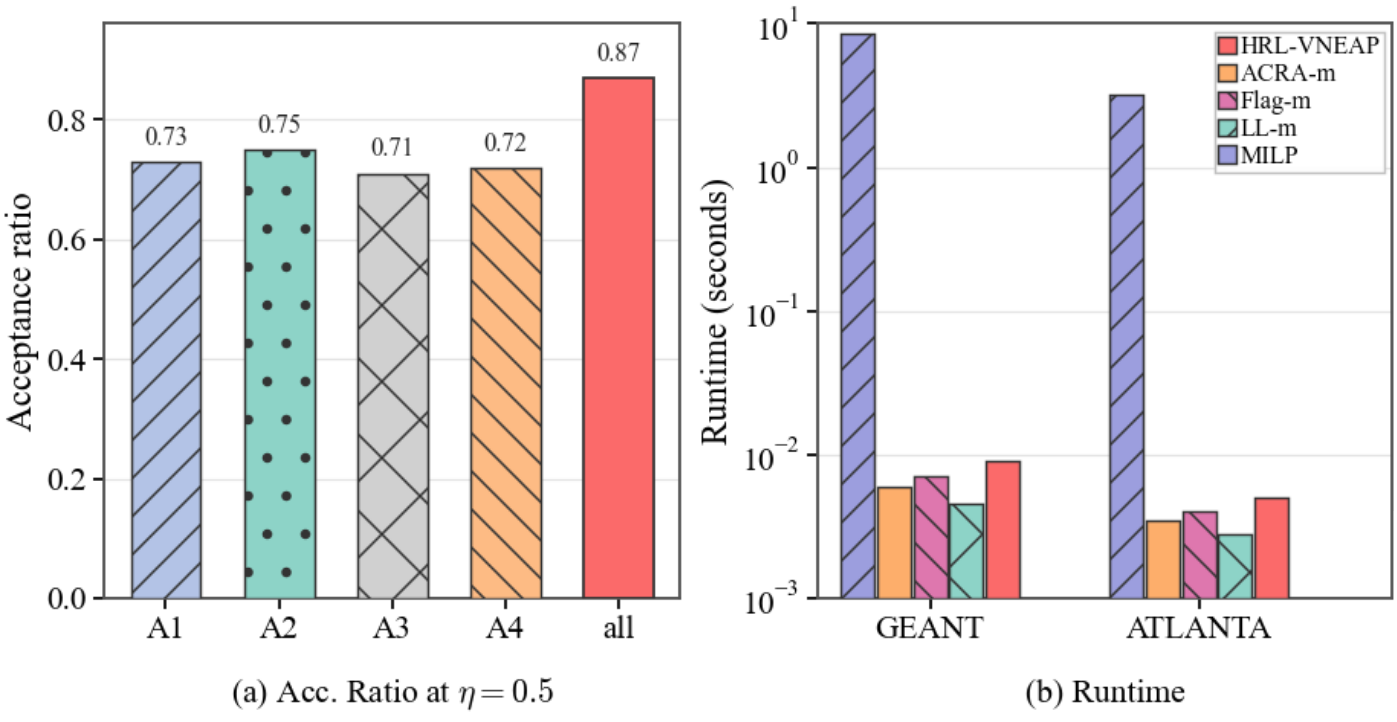}
    \caption{Performance of proposed alternatives and runtime. 
    (a) Acceptance ratio on GEANT at $\eta=0.5$ using each alternative alone or all alternatives jointly. 
    (b) Runtime comparison on GEANT and Atlanta.}
    \label{fig:part3}
\end{figure}

\section{Conclusion}
\label{sec:conclusion}
In this paper, we proposed HRL-VNEAP, a HRL framework for VNEAP. We designed a coupled and efficient policy that enables long-term planning and adaptive decisions. Results across an extensive set of experiments verify the effectiveness of HRL-VNEAP and demonstrate that it outperforms baseline and shows promising results for achieving nearly optimal solutions. Future work includes enriching the alternative-selection space and narrowing the gap to optimality. 


\bibliographystyle{IEEEtran} 
\footnotesize
\bibliography{ref}

\end{document}